# Is AMOC collapse too unstable in climate models?

Sybren Drijfhout[1,2] & Alejandra Sanchez-Franks[3]

[1] University of Southampton, United Kingdom
[2] Royal Netherlands Meteorological Institute, de Bilt, the Netherlands
[3]National Oceanography Centre, Southampton, United Kingdom

**Abstract**
It has been a long-held paradigm that the Atlantic Meridional Overturning Circulation (AMOC) in coupled climate models is too stable, underestimating the possibility of a collapse. A recent analysis of extended simulations to year 2300 in the Coupled Model Intercomparison Project (CMIP)6 (Drijfhout et al., 2025), demonstrated a partial AMOC collapse where the deep circulation completely vanishes. This collapse was simulated by at least one model in all emission scenarios, and it occurred in every model forced by the highest emission scenario when extended to year 2300. Given the modelling community's previous emphasis on stability in climate models, in this perspective, we explore the alternate possible mechanisms by which the AMOC in CMIP6 might have become too unstable. This is motivated by (a) the lack of evidence of downward trends in the AMOC in the observational records, and (b) a subset of CMIP6 models displaying too high climate sensitivity and too strong sensitivity to aerosol removal, both of which could predispose them toward a spurious collapse. For these reasons, it is critical to evaluate the mechanisms by which CMIP6 models simulate AMOC collapse, to be able to develop programs further investigating the trustworthiness of simulated AMOC collapse scenarios.

Keywords:
AMOC
SPG
Collapse
Stability
Model-bias

## 1. Introduction

The possible collapse or shutdown of the AMOC is widely recognised as one of the most severe consequences of continued global warming, due to its possibly abrupt and unstoppable onset, portrayed by the concept of 'tipping point' (Lenton et al., 2019; Armstrong McKay et al., 2021; Lenton et al. (2025). Climate tipping points are often regarded as posing threats of unprecedented magnitude to human civilisation, especially when one tipping point starts a cascade of several other tipping points, causing a domino effect of severe and destabilising changes in the Earth's climate system (Steffen et al., 2018; Wunderling et al., 2021).

As a consequence of this, the possibility of crossing one or more climate tipping points in the coming decades, if we fail halting the emission of fossil fuel-based greenhouse gases, has incited a whole range of international projects (e.g., ClimTip, TipESM, OptimESM, TIPMIP, TiPACCs, ECOTIP, HABITABLE; https://ec.europa.eu) and national research initiatives from which the UK's Advanced Research and Innovation Initiative program – its 'Forecasting Tipping Points' opportunity space – is perhaps the most comprehensive, having funded 27 international projects (https://aria.org.uk/opportunity-spaces/scoping-our-planet/forecasting-tipping-points).

While these scientific projects are delivering a great wealth of new insights on drivers and impacts of various climate tipping points, their perceived risks, and the possible existence early warning indicators of their approach, much of our knowledge is necessarily based on still incomplete and often idealised climate models, since we need to address future events not yet observed, using relatively short-term observational time series that may or may not contain early signs of such future events.

In this perspective we will shortly review our current but still debated understanding of how AMOC tipping may come about in the next decades and what aspects in this chain of events – as currently represented in climate models – may be questioned, especially due to limited resolution, crudely parametrised or unresolved processes, and model biases arising from incomplete interaction between ocean, atmosphere and cryosphere in those models. We suggest specific research initiatives to test whether current climate models misrepresent the various links in the chain of events leading to AMOC tipping, to either enhance our trust in their projections, or where they fail, to guide model improvement towards a more trustworthy projection of future tipping. We also briefly review the observational evidence for AMOC weakening or tipping, and what developments are needed in our monitoring capabilities to extract essential variables needed to develop early warning indicators. We also briefly evaluate the observational databases that are used to benchmark climate models and from which possible early warning indicators for AMOC tipping should be extracted. Where needed we will suggest additional monitoring of essential variables that could facilitate the development of such early earning indicators.

## 2. The AMOC lower limb, deep density gradients and deep convective mixing.

Many CMIP6 models show an AMOC shutdown when simulations are extended to year 2300 (Drijfhout et al., 2025; Figure 1). The simulated shutdown is incomplete, leaving a residual AMOC that no longer consists of an overturning cell associated with North Atlantic Deep Water (NADW) and its return flow. Prior to the AMOC shutdown the models simulate a total collapse of deep convection (Figure 2). This means that net sinking in the SPG completely vanishes, but several decades later than the collapse of deep convection. The sinking branch of the AMOC indeed does not coincide with the sites of deep convection. If the sinking would occur in the open ocean the gyre circulation should be ca 10 times stronger as it observed to satisfy the vorticity budget. Spall and Pickart (2001) argued that sinking must occur where the geostrophic balance no longer holds, i.e. in a boundary current. They linked it to mass conservation resulting from the net northward flow of the AMOC that must sink at the northern boundary of the SPG. One could, however, similarly argue that satisfying mass conservation, the sinking can be linked to AMOC's southward deep return flow away from the northern boundary. This is further motivated by the fact that in CMIP6 models AMOC shutdown always occurs after a deep convection collapse, and the SPG collapse directly impacts the density field at depth. The shutdown of deep convection is illustrated by a collapse of mixed-layer depths in March, when they reach maximum values, from circa 1000 m to less than 200, often even 100 m depth (Figure 2). This collapse of the subpolar mixed-layers occurs much faster (20-30 years) than the shutdown of the NADW-cell of the AMOC which may take 50-100 years (Drijfhout et al. 2025; Angevaare and Drijfhout 2026). For this reason, our main premise is that the chain of events that leads to a shutdown of the AMOC's NADW-cell must be centred around the AMOC's lower limb, the deep density gradients that drive it, and the deep convective mixing that feeds it.

The southward flowing lower limb of AMOC is set by an east-west pressure gradient between ca 1000-2000 m depth. This pressure gradient, mostly determined by heavy, dense water at the western boundary needs to be constantly replenished by dense waters formed by deep convection in the SPG and subsequently transported to the deep western boundary. This feed from deep convection is to balance density loss by mixing with the surrounding waters and southward advection of the existing density anomaly at the western boundary. When the source for this dense water disappears, it takes ca 50 years for the density anomaly and subsequent east-west pressure gradient to be fully eroded, being the lag between convection collapse and shutdown of the AMOC's NADW-cell, and thus also the lag between the collapse of deep convection and the shutdown of net sinking in the SPG.
When the deep, southward branch can no longer be maintained by the deep pressure gradient, the sinking and northward transport of lighter water also stops. This implies that net sinking at subpolar latitudes can no longer be sustained after convection has collapsed, but that there is no direct coupling between the two on timescales less than 50 years. Sinking does not immediately vanish when convection shuts down. Also, deep convection must vanish at all sites for the AMOC's NADW-cell to shut down, including deep convection in the Nordic Seas (Drijfhout et al., 2025). If convection is still present somewhere in the North Atlantic or Nordic Seas the zonal east-west density gradient in the deep ocean can still be maintained, albeit reduced in strength. An illustration of this deep mixing collapse is shown for the MRI-ESM2-0 in both simulations driven by an SSP126 and SSP585 greenhouse gas emission scenario (Figure 2).

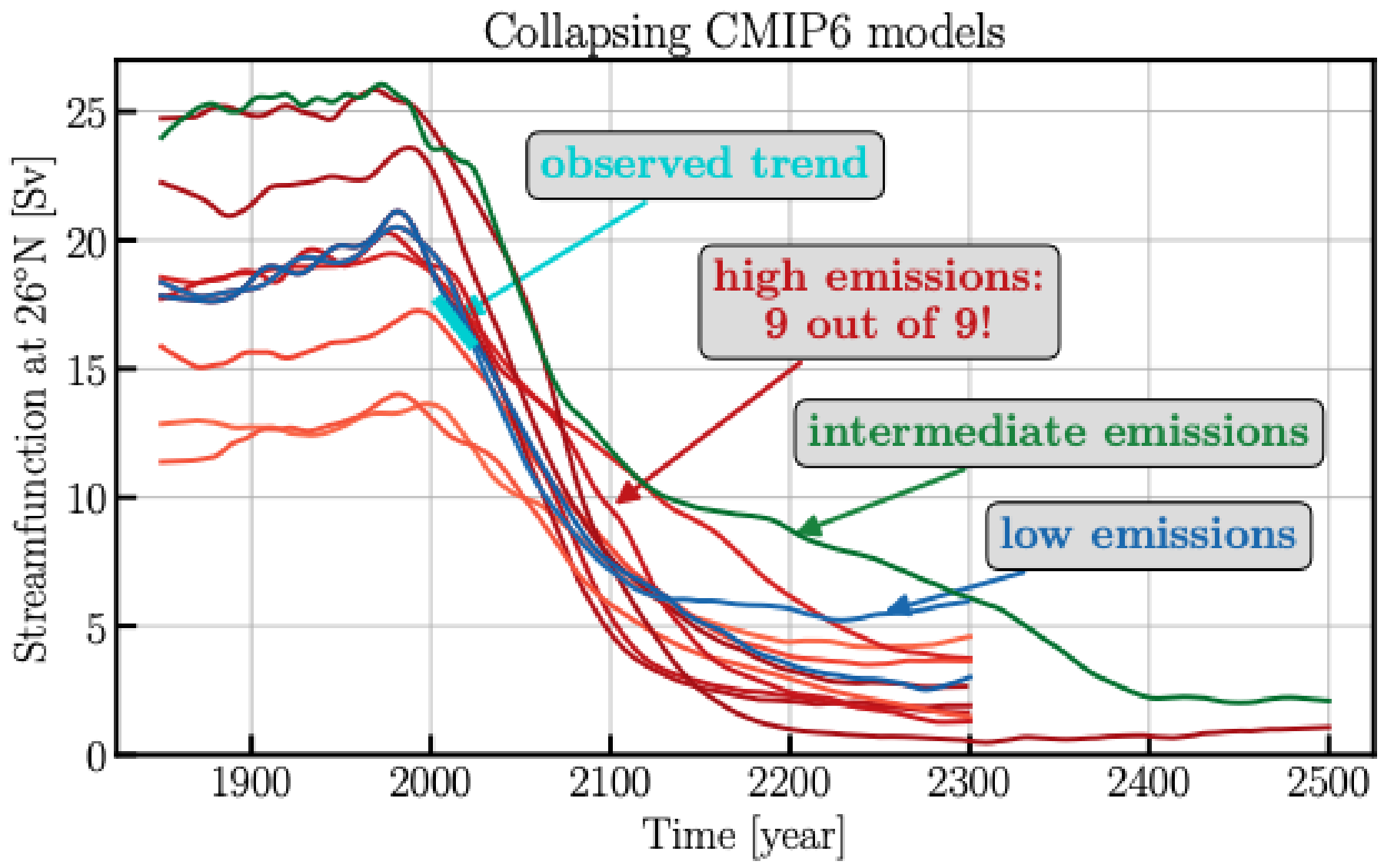

Figure 1. AMOC strength at 26°N in 12 model simulations in which the deep AMOC shuts down. The short cyan line shows the observed (yet insignificant) linear trend of the RAPID observations for 2005-2023 (Volkov et al., 2024).

After the NADW-cell has collapsed the AMOC still features a weaker, cross-hemispheric southward transport of waters subducting in the North Atlantic subtropics, with an associated northward return flow in the surface mixed-layer. This shallow overturning cell no longer transports heat north of the subtropical/subpolar gyre boundary, nor does it release heat from the ocean to the atmosphere at those latitudes (Drijfhout et al., 2025). In other words, for Europe a collapse of the AMOC's NADW-cell has the same climate impact as an AMOC that completely vanished. It should also be emphasized that we currently have no means to assess whether such a weak, residual AMOC is a stable state and whether the transition towards this state can be described by crossing a tipping point, or whether such a state is meta-stable, sometimes described as an edge- or melancholia state from which recovery is still possible (Lohmann and Lucarini, 2024; Lohmann et all., 2025; Moinat et al., 2026). We only have the indication that this state can last for many centuries, making the discussion of its exact stability properties somewhat academic.

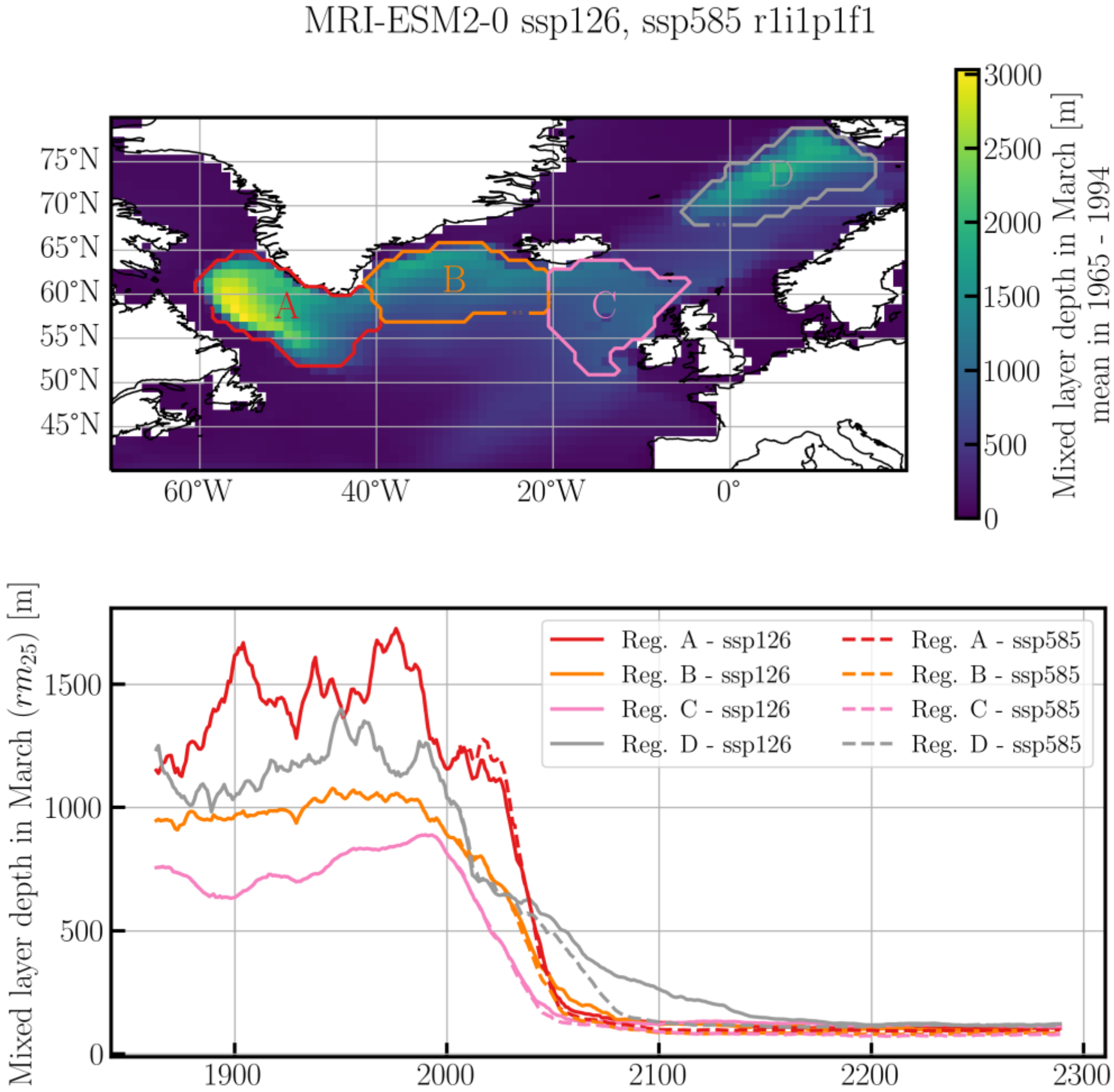

Figure 2. Map of deep mixing and time series of mixing depths in the MRI-ESM2-0. The geographical areas enclosed by a red. orange, pink and grey contour are defined by demanding the March mixed layer between 1965-1995 is deeper than 650m.

## 3. The chain of events leading to AMOC tipping

Because an AMOC shutdown is preceded by a collapse of deep convective mixing in the North Atlantic, the chain of processes leading to this shutdown starts with the convection collapse. Positive feedbacks may ultimately shape the increased stratification of subpolar and Nordic Sea waters, but the initial trigger must come from the forcing, i.e. the greenhouse gas driven changes in surface heat and freshwater fluxes. Several analyses of changes in forcing have elucidated that changes in heat flux over the convective areas are at least an order of magnitude larger than changes in freshwater flux, even if meltwater from the Greenland icesheet is accounted for (Drijfhout et al., 2025; van Westen et al., 2025a, 2025b). Thus, the first step is:

a) Atmospheric warming over the subpolar and Nordic Seas triggers a reduction in deep mixing and less dense waters being formed, further exacerbated by freshwater release from Greenland

The lighter surface waters then lead to lighter watermasses subducting below the mixed layer in late winter and being transported to the western boundary of the deep ocean between ca 1 and 2 km depth. The reduced east-west density/pressure gradient then leads to a weaker southward return flow of the AMOC's lower limb (Ashton-Key et al. (2026). Thus:

b) Weaker deep mixing leads to reduced zonal pressure gradients in the deep ocean

While sinking and deep convection are not directly correlated event (b) emphasizes a lagged response between the AMOC's lower limb and deep convection due to the water masses that subduct or sink below the mixed layer and get entrained into the boundary. Mass continuity then requires a similar lagged response to sinking and the AMOC's upper limb. In other words, the AMOC weakens because its lower limb weakens, and as a result less salty and warm water are transported to the northern North Atlantic. The resulting freshening act as positive feedback increasing the stratification of the water column, while the reduced heat transport counteracts the implied warming imposed by the atmosphere, supporting an increased stratification that is mainly determined by freshening of the surface waters Weijer et al. 2019; VanderBorght et al., 2025). Thus:

c) A weakening AMOC acts to freshen and further stratify thermocline waters in the northern North Atlantic

Reduced deep mixing itself may act as positive feedback on AMOC weakening: with less mixing, surface waters are less well mixed with underlying warm and salty water below. Because that water's temperature signal is more strongly affected by interaction with the atmosphere than its salinity signal, the reduced mixing leads to further freshening of the surface waters (Kuhlbrodt et al., 2001, Born et al., 2012). Thus:

d) Reduced convective mixing acts as a second, positive feedback further freshening the thermocline waters (Figure 3).

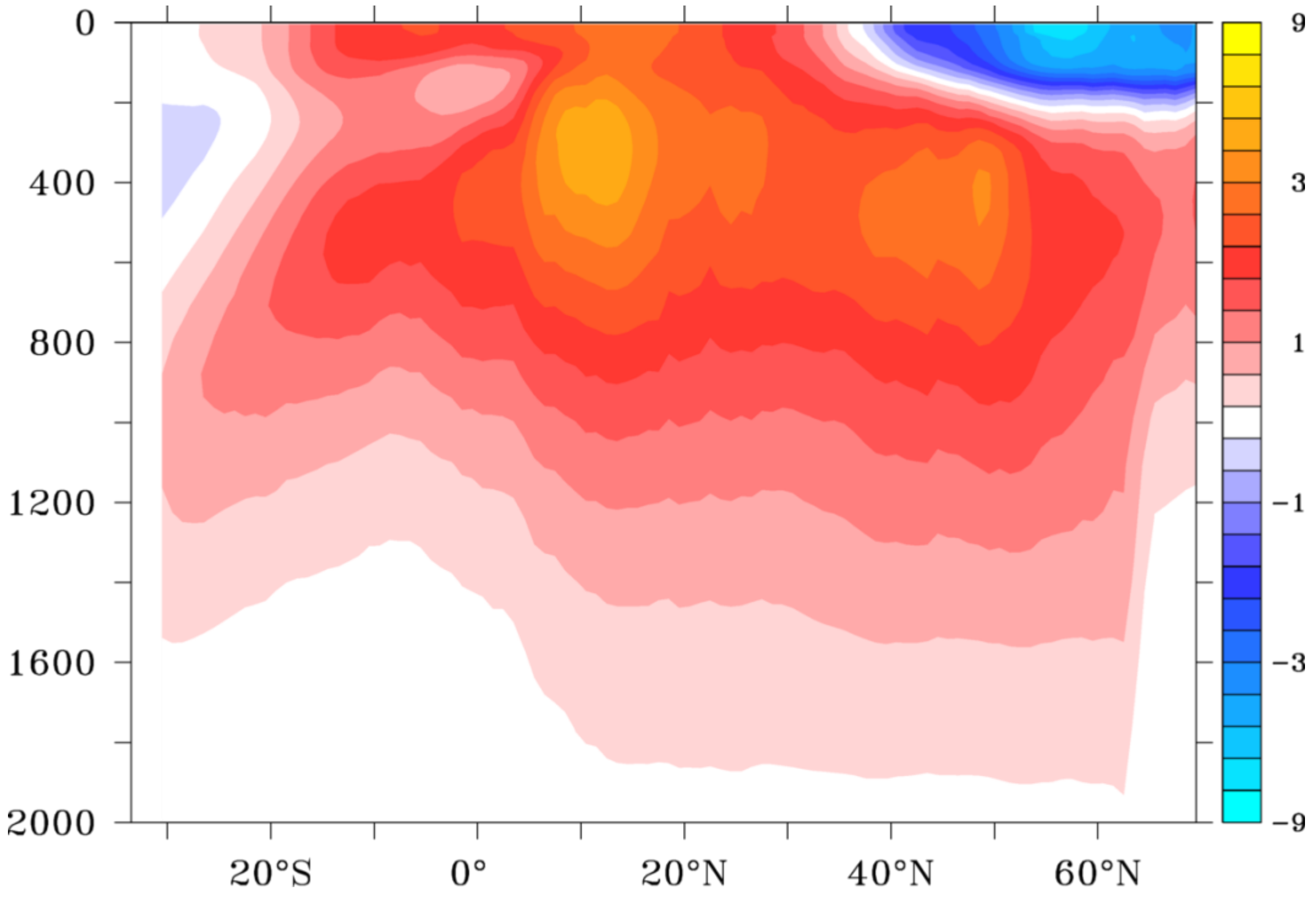

Figure 3. The salinity-change in psu in the Atlantic between 2270-2300 and 1970-2000 as simulated by the CESM-WACCM under ssp585 forcing illustrating the convective mixing feedback by freshening the upper 200 m in the SPG/Nordic Seas, while the rest of the Atlantic was getting saltier.

The decrease in heat loss from ocean to atmosphere may lead to complicated, and not yet well understood atmospheric responses. One response that seems robust is that the atmosphere invokes a seesaw response between North Atlantic subpolar gyre (SPG) and Nordic Seas, initially further decreasing deep convection in the Labrador and Irminger Seas but enhancing it in the Nordic Seas (van den Berk et al., 2021; Iakovleva and Bashmachnikov, 2021). This response may enhance the AMOC's stability by maintaining deep convection in the Nordic Seas. However, while most models do support a more resilient deep mixing in the Nordic Seas relative to the Labrador and Irminger Seas, the convective and advective salt feedbacks overcome the damping of atmospheric effects over the Nordic Seas, limiting the strength of its convection. On a larger scale, the atmospheric response is governed by an increase in atmospheric northward heat transport that partially compensates – the so called Bjerknes feedback – for the reduced oceanic northward heat transport by the AMOC resulting in increased ocean heat uptake in the north, further decreasing ocean heat loss to the atmosphere in the deep-mixing areas (Van der Swaluw et al., 2007; He et al., 2019; Povea-Perez et al., 2023, Garuba et al., 2025). Thus:

e) Enhanced atmospheric northward heat transport acts as Bjerknes feedback further damping heat loss and deep -water formation in the convective areas

The interaction with sea-ice is complex and non-negligible. Initially warming may trigger enhanced melt and freshening in the northern North Atlantic acting as positive feedback, but in response to reduced ocean heat transport, sea ice will start to grow in the area and may reverse its impact on the surface salinity budget through brine rejection during sea-ice growth (Liu and Fedorov 2022, Lee and Liu 2023). This negative feedback appears to start working only after the deep mixing has collapsed, so it does not act as a damping factor in the whole chain of events. This negative feedback could act to limit the convection collapse or support some other mechanism to restart deep convection; on its own, however, it appears too weak to prevent the deep mixing convection collapse. Thus:

f) Increased sea-ice formation, in response to the reduced ocean heat transport, acting to salinify the upper-layer waters in the northern North Atlantic only become active after those waters have been freshened enough to collapse deep convection and its impact is too weak to prevent this collapse.

**4. What are the weak links in this chain of events and what research topics should be prioritised?**

Our main concern given the relatively large set of models showing an AMOC shutdown is that the CMIP6 models may lack stability, contrary to previous studies suggesting it is too stable (Valdes 2011; Mecking et al. 2017; Rahmstorf 2024, Van Westen et al. 2024). This lack of stability has been fuelled by the lack of significant trends in the observations (McCarthy et al. 2023, Volkov et al., 2024). A second argument for a possible lack of stability is the too high climate sensitivity combined with a too strong aerosol effect in the last generation of climate models, making them too sensitive to the applied forcing (Robson et al., 2022). This is indeed a concern, and if true, would certainly bias the timing of a collapse of deep convective mixing and AMOC shutdown. Therefore, we want to emphasize here, which miss- or incompletely represented processes might cause CMIP6 models being too unstable, without suggesting they probably are, because if they are not too unstable, the implications of simulated AMOC and SPG collapse would indeed be very worrying if not threatening. We suggest focusing on whether low-resolution models miss or misrepresent physical processes that would fundamentally change the stability properties of convective mixing and AMOC shutdown. Therefore, we suggest to carefully check whether not resolving processes or dynamical features due to resolution limits affects the separate links in the chain of events (a – f) described above.

a) Warming triggers reduced deep mixing, and less dense waters are being formed (Fig. 4).

- Stratification bias impacts this link. Is the trigger too easily accomplished in low-resolution coupled models? This can be investigated by checking the stratification bias versus AMOC decline and by using box models of convective instability tuned to a set of CMIP6 and high-resolution coupled- and forced models, and to observations; and estimating whether low-resolution CMIP6 models overestimate the trigger.
- Western Boundary Currents heat up three times faster as the interior ocean due to eddy fluxes concentrating the heat in the currents instead of transporting across, implying different behaviour from the Gent and McWilliams parameterization (Wu et al., 2012). Resolved eddy transports behaving different from standard parameterizations also happens in the Pacific sector of the Southern Ocean (Kang et al., 2026). Can something similar occur around convective areas in the Northern Atlantic? It should therefore be analysed whether eddies trap heat in the surface waters there, resisting the effect of warming on stratifying the water column in high-resolution coupled- and forced ocean models relative to low-resolution CMIP6 models.

b) Link between deep convection and density anomalies at the western boundary of the deep North Atlantic is too weak

- This can be checked by performing Lagrangian backtracking in CMIP6 and high-resolution coupled- and forced ocean models to check pathways and advection timescales and the role of other mixing processes in establishing the dense waters at the western boundary relative to the waters at the eastern boundary, driving, through the associated zonal pressure gradient the AMOC's lower limb and whether these are modified by eddy fluxes. In particular, it should be investigated whether eddies destroy such a direct link by creating more wobbly pathways with a lot of recirculation, enhancing the role of interior mixing versus deep convection.

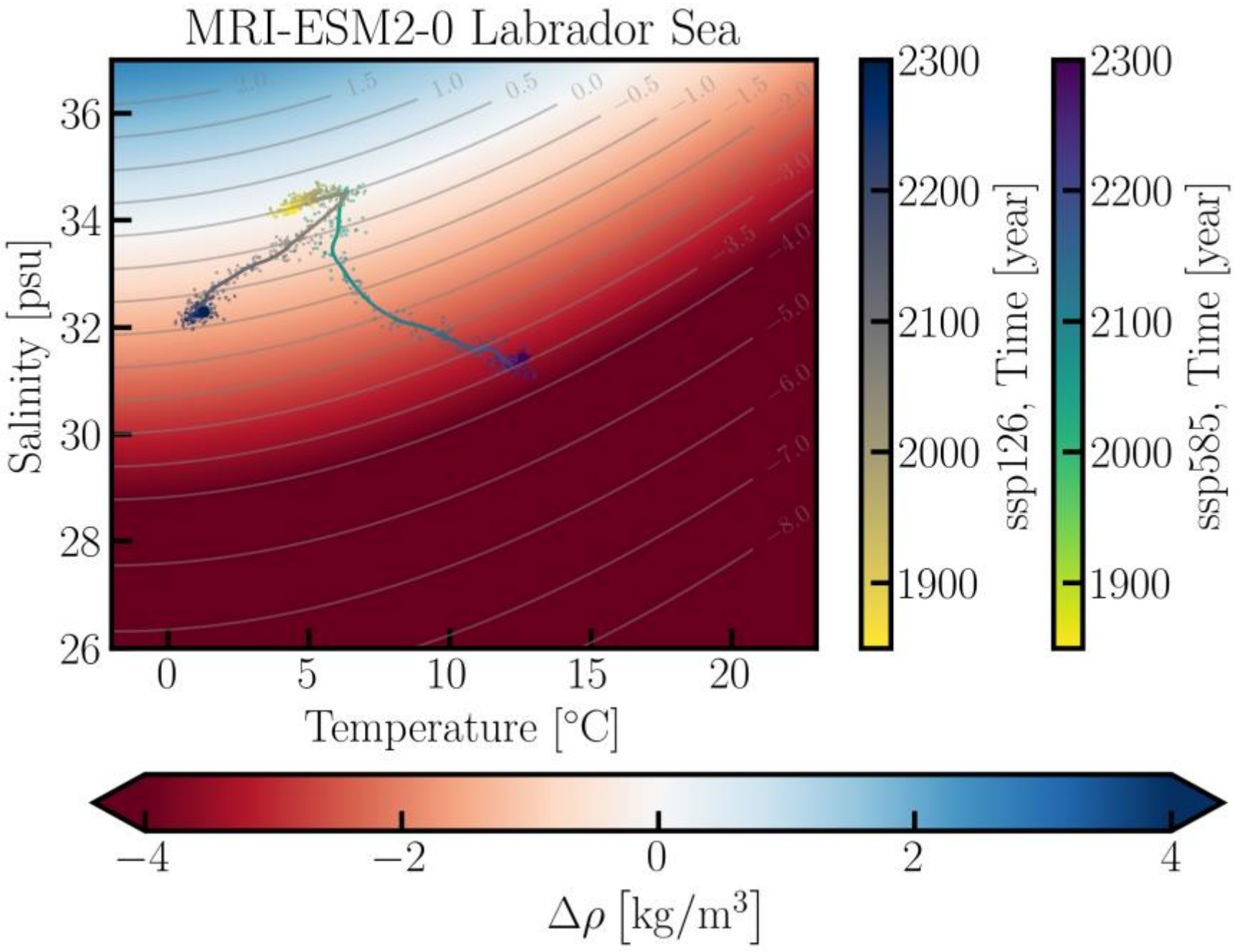

Figure 4. The trajectory in surface watermass (temperature, salinity space) in the Labrador Sea deep-mixing region as a function of time, starting with yellow dots and ending with blue dots as simulated by the MRI-ESM2-0 forced by an ssp126 (left curve) and ssp585 scenario (right curve), illustrating the gradual increase in buoyancy by first temperature increase, followed later by salinity decrease.

c) The advective salt feedback is too strong

- The salt feedback is important to describe the interaction between AMOC and deep convection in the North Atlantic, but in transient runs the feedback operating at the boundary between subtropical and subpolar gyre becomes decisive. The question then becomes whether this feedback is too strong in CMIP6 models. For the multi-model mean ensemble, the answer is no (Mecking et al., 2017; van Westen and Dijkstra 2024), but we don't know whether models in which the AMOC collapses feature stronger feedback than models in which the AMOC does not collapse and whether these two groups feature a stronger, respectively weaker feedback than observed. This should be checked. Also, it might be possible that the vertical profile of the feedback matters and whether this is connected to temperature/salinity biases in the vertical in the deep mixing areas. This should be checked as well.
- In addition, we suggest calculating horizontal pathways of salinity-decrease as a function of AMOC weakening in CMIP6-, coupled, forced high-resolution models, and the ECCO state estimate product to establish whether eddies and thinner, swifter and more turbulent currents modify pathways and change the pathways of the freshwater fluxes into the convective regions, making the same salt feedback in terms of strength more or less effective in affecting the convective areas.

d) The convective feedback affects the stratification too much

- A reduction in deep convection will affect the stratification in the column and on its turn affect the convection further. The strength of this feedback (and even sign) crucially depends on the temperature and salinity signature of the waters in a vertical column and biases in stratification directly translate into biases in convective feedback (Sgubin et al., 2017; Swingedouw et al., 2021). A careful analysis of this should be carried out, comparing models in which the AMOC collapses versus models where it does not collapse and these should be benchmarked by observations.
- Also, it could be that the stratification of models showing a convective mixing collapse is already stronger than observed in their historical rums and whether mixing collapses too early after too little change in forcing.

e) The simulated Bjerknes feedback is too strong

The atmospheric response to changed temperatures and heat fluxes originating from ocean circulation changes generally reinforces convection and AMOC decline. This is mainly associated with the so called Bjerknes feedback implying changes in ocean heat transport are compensated by changes in atmospheric heat transport. This feedback, however, is demonstrated for natural fluctuations in a climate that is in equilibrium. When net ocean heat uptake is no longer zero, this feedback should be less effective. Also, cloud feedbacks might play an important role in changes in surface heat fluxes as has been demonstrated for the development of the so-called warming hole in the North Atlantic subpolar gyre (Li et al., 2024).

- The amount of Bjerknes compensation should be checked in CMIP6 and high-resolution coupled models to assess whether this feedback is too strong in models in which the AMOC collapses and weaker in models where it does not. This can be compared with the operation of the Bjerknes feedback in the ECCO state estimate product.
- Also, it should be checked which cloud feedbacks operate when convection and outgoing heat flux reduce and whether this is different between models showing tipping versus models that do not show a collapse in deep mixing for a given forcing scenario. Again, this should be compared with high-resolution coupled models and observational and reanalysis products.

f) The interaction between ocean and declining sea-ice cover is incorrect

- Reduced ocean heat transport may counteract further sea-ice melt imposed by global warming by reversing or damping the temperature trend associated with global warming over the northern North Atlantic. The subsequent modification of the sea-ice seasonal growth and decline cycle affects the freshwater balance in the northern North Atlantic and this process could counteract or delay the convection collapse. Whether this interaction is correctly modelled should be subject of further research.
- The interaction between reduced heat transport by the AMOC and the seasonal cycle of sea-ice, modifying amounts and sites where sea-ice grows injecting salinity into the subsurface waters and where it melts adding freshwater to the surface should be investigated in CMIP6, coupled and forced high-resolution models and reanalysis products.

## 5) Observational Constraints and the Challenge of Detecting AMOC Tipping

a) The state of the AMOC from the observations

To date, modern oceanographic observations have not documented a tipping event in the AMOC. Instead, a primary objective of current research is to determine whether a long-term weakening trend exists. Direct estimates from OSNAP programme in the Subpolar North Atlantic (56°N; Lozier et al., 2017), available only since 2014, are too short to robustly resolve multi-decadal variability. At lower latitudes, the RAPID array has provided continuous AMOC measurements since 2004 (26°N; McCarthy et al., 2025). Within this record, the Florida Current—measured via a submarine cable since 1982 —has remained broadly stable over approximately four decades, while the full AMOC record exhibits a weak negative trend that is only marginally statistically significant (e.g. Volkov et al., 2024). Other observational studies similarly report no statistically significant long-term trend in AMOC strength or western boundary transports (Moat et al., 2020; Worthington et al., 2021; Le Bras et al., 2023; Rossby et al., 2014; Fu et al., 2020). A key limitation is the relatively short duration of observational records. Detecting climate-scale trends in the AMOC likely requires continuous records of 30–70 years, whereas the RAPID time series currently spans just over two decades (Lobelle et al., 2020). This limits confidence in distinguishing long-term change from internal variability. Further, similar to the climate models, our ability to observe convection itself is similarly resolution-limited: gridded observational products are typically too coarse (~1°) to resolve the mesoscale eddies that govern convective heat and salt fluxes and/or resolve the rapid transition of the mixed-layer depth around its critical March peak. This observational resolution gap mirrors the low-resolution parameterisation issues raised for the models in Section 4a.

The key message is that current or modern ocean observations cannot yet refute or confirm evidence for AMOC weakening as they are not long enough to resolve climatic timescales, differentiating between internal variability and climate change. However, there is strong observational evidence of unprecedented change in the North Atlantic: The deep western boundary current component of the AMOC at 26°N has declined over the past two decades (Xing et al., 2026), and the subpolar North Atlantic "warming hole" — a region of anomalous cooling south of Greenland (Duchez et al., 2016) — has been linked to reduced overturning strength in models (Drijfhout et al., 2012). A pronounced freshening event between 2012 and 2016, the largest in over a century, has similarly been associated with changes in circulation and deep-water formation (Holliday et al., 2020). Direct evidence of change in convection itself comes from the Labrador Sea, where episodic reductions and temporary shutdowns of deep convection in 2021–2023 have been linked to atmospheric forcing and freshwater anomalies (Yashayaev et al., 2024), and from a 70-year hydrographic record in the Greenland Sea showing a reduction in density and the formation of a new intermediate water mass, with implications for the AMOC (Strehl et al., 2024).

Determining how and if this evidence is linked to AMOC weakening or collapse remains an open challenge.

b) Which early warning signals for the AMOC have been proposed, and do they work in the real world?

Recent publications on approaching tipping points for the AMOC assume this circulation can be approximated by a single time series – a spatial average of temperature and salinity anomalies over a region of the SPG, used a simplified proxy for the AMOC (Boers, 2021; Ditlevsen and Ditlevsen, 2023). In a follow-up study, however, it was argued that such methods are too uncertain to give robust estimates (Ben-Yami et al. 2024). Timeseries of the AMOC itself are too short, to find EWS based on statistical methods focusing on Critical

Slowing Down (CSD; Scheffer et al., 2009), hence the attempt to find CSD in AMOC fingerprints of temperature that has longer timeseries. The chance that CSD can be applied if AMOC monitoring would be extended in the future is however small, as the AMOC is possibly prone to noise-induced tipping and the relevance of CSD in circumstances of fast (relative to the timescales of AMOC internal variability) climate change is uncertain. As a result, CMIP6 models do not show CSD of the AMOC in their historical simulations (Dijkstra et al., 2026). Sometimes, knowledge of the edge-state of the AMOC is advocated as EWS, as demonstrated in models of intermediate complexity (Lohmann et al., 2025) and assuming it can be monitored by measuring the three-dimensional temperature and salinity field of the ocean, sometimes collapsed to vertical and domain averages, but it is unclear how this edge-state would look like in the real world or even in CMIP6 models.

## 6) Recommendations and conclusions

Similarly, current numerical models remain too coarse to resolve mesoscale dynamics, while their hydrostatic parameterisations provide only a crude representation of convective mixing. The adequacy of these parameterisations for simulating and forecasting changes in convective activity therefore remains uncertain, although it should be noted that these simple schemes might be able to correctly simulate changes in watermass properties on annual to decadal timescales relevant for monitoring and forecasting purposes. To improve the forecast reliability of convective activity changes, however, also model bias needs to be corrected or removed, especially when it affects the stratification in convective areas, as the moment of a convection collapse crucially depends on stratification changing such that the surface waters, even cooling to the freezing point, still remain lighter than the density at depth. This criterion of a critical surface density, more specifically critical salinity as the surface temperature is allowed to decrease to -1.9°C, has been used to forecast changes in convective activity (Almeida and Swingedouw 2026). Since the moment of passing the critical salinity threshold generally coincides with the occurrence of a convection collapse and thus is unfit to act as predecessor of such a collapse, this coincidence underscores the essential role of stratification in the SPG when trying to forecast a convection collapse, as well as the relative minor role of atmospheric weather once the critical threshold in stratification has been reached.

Our main message thus is that the way forward should consist of combined and synergetic observation and modelling programs where the veracity of the models is improved and the gaps in the observations removed to construct a working monitoring program of the SPG and AMOC. We would prioritise within the chain of events the links between weakening deep convection and AMOC decline with primarily focus on how well stratification in the early 21st century matches observations and whether the convective (salinity) feedbacks are modelled trustworthy. This implies both checking and evaluating density and salinity profiles in the deep convective regions. A second bullet on the list is then associated with the role of mesoscale eddies and improved overflows (parameterisations) to improve the representation of denser Lower NADW south of Denmark Strait, which is presently absent in almost every climate model leading to too light bottom waters in the areas of deep convective mixing.

**Data Availability.** No new data were generated for this issue. All figures and analysis are based on CMIP6 data available from ESGF nodes

**Acknowledgments.** We acknowledge the World Climate Research Programme's Working Group on Coupled Modelling, which is responsible for CMIP, and we thank the climate modelling groups for producing and making available their model output. For CMIP the U.S.

Department of Energy's Program for Climate Model Diagnosis and Intercomparison provides coordinating support and led development of software infrastructure in partnership with the Global Organization for Earth System Science Portals. We acknowledge the World Climate Research Programme, which, through its Working Group on Coupled Modelling, coordinated and promoted CMIP6. We thank the climate modelling groups for producing and making available their model output, the Earth System Grid Federation (ESGF) for archiving the data and providing access, and the multiple funding agencies who support CMIP6 and ESGF. The CMIP6 data was accessed and analysed using the UK super-data-cluster JASMIN and Netherlands super-computer-cluster SURFsara.
SD received funding from OptimESM and TipESM. TipESM "Exploring Tipping Points and Their Impacts Using Earth System Models" is funded by the European Union, Grant Agreement number: 101137673." OptimESM "Optimal High Resolution Earth System Models for Exploring Future Climate Change" is funded by the European Union, Grant Agreement number:101081193. SD received funding from ISOTIPIC. "ISOTIPIC" Interacting ice Sheet and Ocean Tipping - Indicators, Processes, Impacts and Challenges" is funded by the National Environmental Research Council, UK, under Grant Number: NE/Y503320/1. SD and ASF received funding from SORTED. "Subpolar Gyre observations, models and ai to resolve tipping points and provide early warning detection" is funded by the Advanced Research and Invention Agency of the UK under SCOP-PR01-P020 Standard Grant Agreement.